\documentclass[amsmath,amssymb]{revtex4}
\usepackage{times}
\usepackage{graphicx}
\usepackage{amsmath}

\def\s{\sigma}
\def\S{\Sigma}
\def\w{\omega}
\def\e{\epsilon}
\def\d{\delta}
\def\G{\Gamma}

\begin{document}
\title{Current and Noise in a FM/quantum dot/FM System}
\author{F. M. Souza,$^{1,2}$ J. C. Egues$^{2,3}$ and A. P. Jauho$^1$}
\address{$^1$ Mikroelektronik Centret, Danmarks Tekniske Universitet,
DK-2800 Kgs. Lyngby, Denmark\\
$^2$ Departamento de F{\'\i}sica e Inform{\'a}tica, Instituto de
F{\'\i}sica de S{\~a}o Carlos,\\ Universidade de S{\~a}o Paulo,
13560-970 S{\~a}o Carlos, S{\~a}o
Paulo, Brazil.\\
$^3$ Department of Physics and Astronomy, University of Basel,\\
Klingelbergstrasse 82, CH-4056 Basel, Switzerland}
\date{\today}
\begin{abstract}
Using the Keldysh nonequilibrium technique we calculate current,
noise and Fano factor in a ferromagnetic(FM)-quantum
dot-ferromagnetic(FM) system with Coulomb interaction and
spin-flip scattering in the dot. The lead polarizations are
considered in both parallel P and antiparallel AP alignments. We
show that spin-flip can increase both AP-current and AP-noise,
while the P-current and P-noise are almost insensible to it. This
fact leads to a suppression of the tunnelling magnetoresistance
with increasing spin-flip rate.
\end{abstract}
\maketitle

\section{Introduction}
The emerging field of spintronics \cite{dda02}-\cite{gap98}, where
the electron spin and charge are used to design new devices, has
led to fascinating and novel ideas such as spin filters
\cite{rf99}-\cite{tk02}, spin field effect transistors
\cite{sd90}, and has offered many proposals for solid state
quantum computing \cite{dpv95}. For example, quantum dot systems
are useful in the control of the electron spin and are suitable to
create quantum bits relevant for quantum gate operations
\cite{hae01}.

The study of nonequilibrium transport properties of spintronic
devices is of great importance to understand basic physical
phenomena and to predict new functionalities. Calculation of the
current, for example, can give the conductance/resistance of a
system and its dependence on magnetic field, Coulomb interaction,
spin-flip and so on. On the other hand, current fluctuations, due
to the granularity of the charge (shot noise \cite{ymb00}), are
also relevant because their measurements can provide additional
information not contained in the average current \cite{clk97}.

Here we apply the Keldysh nonequilibrium technique \cite{lvk65} to
calculate current and its fluctuations (noise) in a quantum dot
coupled to two ferromagnetic leads as a function of the applied
voltage for parallel and antiparallel lead-polarization
alignments. We include Coulomb interaction in the Hartree-Fock
approximation as well as spin-flip in the dot. We show that
spin-flip makes the alignment of the lead polarizations less
important; both P and AP results coincide for large enough
spin-flip rates. This fact gives rise to a reduction of both Fano
factor and tunnelling magnetoresistance (TMR) as we show here.

Our paper is organized as follows. In Sec. 2 we describe the
system and present its Hamiltonian. In Sec. 3 we apply the Keldysh
technique to determine current and noise in our system. In Sec. 4
we discuss our results for current and noise and Sec. 5 gives our
conclusions.

\section{System}

\begin{figure}
\begin{center}
\includegraphics{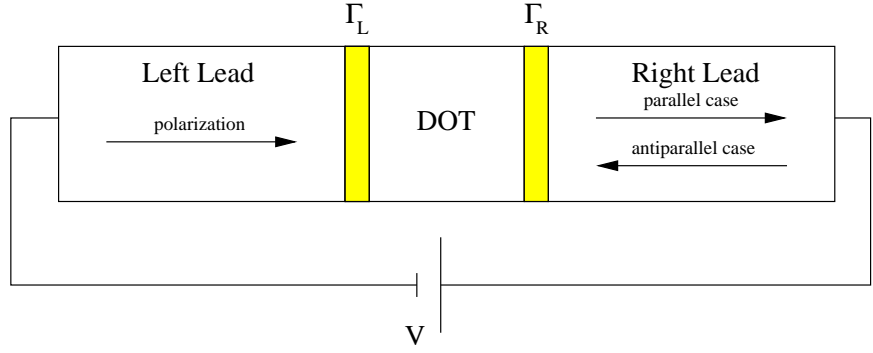}
\end{center}
\caption{Schematic of the system. It is composed of two
ferromagnetic leads and a quantum dot as a spacer. The electrons
are allowed to tunnel through the left and right barriers (with
tunnelling rate $\G_L$ and $\G_R$, respectively) in order to
generate a tunnelling current when a voltage $V$ is applied. The
left lead has a fixed polarization (hard side) while the right
lead switches its polarization from parallel (up arrow) to
antiparallel (down arrow) alignment. This polarization rotation
changes the majority/minority spin population, leading to a
variation in the resistance of the system, which is reflected in
both current and noise.} \label{fig1}
\end{figure}

Our system is composed of two ferromagnetic leads coupled to a
quantum dot via tunnelling barriers (Fig.\ref{fig1}). While the
left lead has a fixed polarization (hard lead), the right one can
have its polarization switched from parallel P to antiparallel AP
alignment (soft lead). This polarization rotation
(P$\rightarrow$AP) changes the transport properties of the system
\cite{commentGMRPrinz}. This effect is included in our approach.

We model the system with the Hamiltonian $H=H_L+H_R+H_D+H_T$,
where $H_{L(R)}$ is the left (right) lead Hamiltonian, $H_D$
describes the dot and $H_T$ gives the coupling between leads and
dot. In our model, Coulomb interaction and spin-flip are
restricted to the dot, while the electrons in the leads are free.
The leads are assumed to be in thermal equilibrium with chemical
potential $\mu_L$ and $\mu_R$ for the left and the right leads,
respectively. When a voltage $V$ is applied across the system, the
chemical potentials differ by $\mu_L-\mu_R=eV$, where $e$ is the
electron charge. This difference drives the system out of
equilibrium, thus giving rise to current and noise. More
explicitly, the Hamiltonian of the lead $\eta$ ($\eta=L,R$) is
\begin{equation}
  H_\eta=\sum_{k\s}\e_{k\eta\s}c_{k\eta\s}^\dagger c_{k\eta\s},
\label{HLHR}
\end{equation}
where $c_{k\eta\s}$ ($c_{k\eta\s}^\dagger$) destroys (creates) an
electron into the lead $\eta$ with wave vector $k$ and spin $\s$.
The electron energy $\e_{k\eta\s}$ depends on $\eta$ and the spin
component $\s$ because of the applied voltage and the band
spin-split, respectively.

The Hamiltonian of the dot is
\begin{equation}
  H_D=\sum_{\s}\e_0 d_\s^\dagger d_\s+Un_{\uparrow}n_{\downarrow}+R(d_{\uparrow}^\dagger
  d_{\downarrow}+d_{\downarrow}^\dagger d_{\uparrow}),
\label{HD}
\end{equation}
where $d_\s$ ($d_\s^\dagger$) destroys (creates) an electron in
the dot with spin $\s$ and the energy $\e_0$ is spin independent
\cite{wr01}, \cite{pz02}. In addition, we assume we have a small
enough dot in order to have only one active level $\e_0$. In the
presence of a voltage the level shifts by
$\e_0=\e_{d}-\frac{eV}{2}$, where $\e_{d}$ is the dot level for
zero bias (for numerical convenience we use $\e_{d}=\frac{U}{2}$).
This assumption does not take into account charge accumulation in
the dot, which tends to wash out this linear drop. A more
sophisticated approach, which includes charging effects in a
self-consistent way, will be discussed elsewhere \cite{fms02}. In
Eq.(\ref{HD}) the Coulomb interaction is taken into account via
the Hubbard term with a correlation parameter $U>0$ and spin-flip
scattering is described by the last term, where $R$ is the
spin-flip scattering amplitude.

The tunnelling Hamiltonian is
\begin{equation}
  H_T=\sum_{k\eta\s}(t_{k\eta\s}^* d_\s^\dagger c_{k\eta\s}+t_{k\eta\s}c_{ k\eta\s}^\dagger
  d_\s),
\label{HT}
\end{equation}
where $t_{k\eta\s}$ couples an electronic state in lead $\eta$ to
one in the dot. We consider a spin conserving tunnelling; the
spin-flip process is assumed to be confined in the dot. In the
nonequilibrium Green function technique $H_T$ is the
nonequilibrium part of the Hamiltonian because it couples contacts
with different chemical potential (if $eV\neq 0$), thus allowing
for charge flow. Next we apply the Keldysh technique \cite{hh96}
to determine the average current and the noise.

\section{Current and Noise}

{\it Current.} The average current from the left contact into the
dot is defined as $I_L=-e\langle\dot{N}_L\rangle$, where
$N_L=\sum_{k\s}c_{kL\s}^\dagger c_{kL\s}$ is the number operator
for lead $L$. To find the time evolution of the occupation-number
operator, we use the Heisenberg equation $\dot{N}_L=i[H,N_L]$. The
only term of the Hamiltonian which does not commute with $N_L$ is
$H_T$. Using Eq.(\ref{HT}) we obtain
\begin{equation}
  I_L=\frac{2e}{\hbar}{\mathrm{Re}}\sum_{k\s}t_{kL\s}i\langle
  c_{kL\s}^\dagger (t) d_\s (t)\rangle.
\label{ILsecCurrent}
\end{equation}
To avoid further complications in the analysis due to the
spin-flip term, we perform a canonical transformation \cite{pz02},
\begin{equation}
 \begin{pmatrix}
   d_{\uparrow} \\
   d_{\downarrow} \
 \end{pmatrix}=\frac{1}{\sqrt{2}}
 \begin{pmatrix}
   1 & 1 \\
   -1 & 1 \
 \end{pmatrix}
 \begin{pmatrix}
   d_{1} \\
   d_{2} \
 \end{pmatrix},
\label{cannonical}
\end{equation}
in terms of which Eq.(\ref{ILsecCurrent}) becomes
\begin{equation}\label{}
  I_L=\frac{2e}{\hbar}\frac{1}{\sqrt{2}}\mathrm{Re}\sum_{k}
  \mathrm{Tr}\{
  \begin{pmatrix}
    t_{kL\uparrow} & t_{kL\uparrow} \\
    -t_{kL\downarrow} & t_{kL\downarrow} \
  \end{pmatrix}
  \begin{pmatrix}
    G_{1,kL\uparrow}^< & G_{1,kL\downarrow}^< \\
    G_{2,kL\uparrow}^< & G_{2,kL\downarrow}^< \
  \end{pmatrix}\},
\end{equation}
where $G_{i,k\eta\s}^<(t,t)=i\langle c_{k\eta\s}^\dagger (t) d_i
(t)\rangle$. Applying the Keldysh technique as described in
\cite{hh96} we find
\begin{equation}
  I_\eta=\frac{2e}{\hbar}\mathrm{Re}\int dt_2 \mathrm{Tr}
  \{{\mathbf{G}}^r(t,t_2){\mathbf{\Sigma}}^{\eta <}(t_2,t)+{\mathbf{G}}^<(t,t_2){\mathbf{\Sigma}}^{\eta a}(t_2,t)\},
\label{Ieta}
\end{equation}
where $\mathbf{G}^r$ and $\mathbf{G}^<$ are the nonequilibrium dot
Green functions, with elements $G_{ij}^<(t,t_2)=i\langle
d_j^\dagger (t_2) d_i(t)\rangle$ and
$G_{ij}^r(t,t_2)=-i\theta(t-t_2)\langle\{d_i(t),d_j^\dagger(t_2)\}\rangle$.
Here the averages are taken over the initial ($t=-\infty$)
equilibrium density matrix \cite{commentRHO}. The lesser
(retarded, advanced) self-energy is given by
\begin{equation}
  \mathbf{\S}^{L<(r,a)}(t_2,t)=\frac{1}{2}\sum_{k}|t_{kL}^2|\begin{pmatrix}
    g_{kL\uparrow}^{<(r,a)}(t_2,t)+g_{kL\downarrow}^{<(r,a)}(t_2,t) & g_{kL\uparrow}^{<(r,a)}(t_2,t)-g_{kL\downarrow}^{<(r,a)}(t_2,t) \\
    g_{kL\uparrow}^{<(r,a)}(t_2,t)-g_{kL\downarrow}^{<(r,a)}(t_2,t) & g_{kL\uparrow}^{<(r,a)}(t_2,t)+g_{kL\downarrow}^{<(r,a)}(t_2,t) \
  \end{pmatrix},
\label{SelfL}
\end{equation}
where $g_{kL\s}^{<(r,a)}$ is the lesser (retarded, advanced)
uncoupled Green function for lead $L$. These are defined as
$g_{kL\s}^< (t_2,t) =i\langle \widetilde{c}_{kL\s}^\dagger (t)
\widetilde{c}_{kL\s}(t_2)\rangle$,
$g_{kL\s}^r(t_2,t)=-i\theta(t_2-t)\langle
\{\widetilde{c}_{kL\s}(t_2),\widetilde{c}_{kL\s}^\dagger(t)\}\rangle$
and $g_{kL\s}^a(t_2,t)=i\theta(t-t_2)\langle
\{\widetilde{c}_{kL\s}(t_2),\widetilde{c}_{kL\s}^\dagger(t)\}\rangle$,
where the tilde denotes that the operator is in the interaction
picture; its time evolution is governed entirely by
Eq.(\ref{HLHR}). In Eq.(\ref{SelfL}) we assume a spin-independent
amplitude $t_{kL}$ for simplicity.

For a time-independent Hamiltonian the Fourier transform of
Eq.(\ref{Ieta}) yields
\begin{equation}
  I_L=\frac{e}{\hbar}\int \frac{d\w}{2\pi}
  \mathrm{Tr}\{\mathbf{\S}^{L<}(\w)[\mathbf{G}^r(\w)-\mathbf{G}^a(\w)]-\mathbf{G}^<(\w)[\mathbf{\S}^{Lr}(\w)-\mathbf{\S}^{La}(\w)]\},
\label{ILfinal}
\end{equation}
where $\mathbf{\S}^{\eta <}(\w)$ and the difference
$\mathbf{\S}^{\eta r}(\w)-\mathbf{\S}^{\eta a}(\w)$ are calculated
using the expressions
$g^{r,a}_{k\eta\s}(\w)=\frac{1}{\w-\e_{k\eta\s}\pm i\eta}$ and
$g_{k\eta\s}^<(\w)=2\pi i n_\eta (\w) \d(\w-\e_{k\eta\s})$,
$n_\eta(\w)$ is the Fermi distribution function of the lead
$\eta$. We find $\mathbf{\S}^{\eta <}=in_{\eta}\mathbf{\G}^{\eta}$
and $\mathbf{\S}^{\eta r}-\mathbf{\S}^{\eta
a}=-i\mathbf{\G}^\eta$, with
\begin{equation}\label{}
    \mathbf{\G}^\eta=\frac{1}{2}
  \begin{pmatrix}
    \G_{\uparrow}^\eta+\G_\downarrow^\eta & \G_\uparrow^\eta-\G_\downarrow^\eta \\
    \G_\uparrow^\eta-\G_\downarrow^\eta & \G_\uparrow^\eta+\G_\downarrow^\eta \
  \end{pmatrix},
\end{equation}
where $\G_{\s}^\eta=2\pi\sum_{k}|t_{k\eta}|^2\d(\w-\e_{k\eta\s})$.

Accounting for Coulomb interaction in the Hartree-Fock
approximation, we can write down a matrix Dyson equation for the
retarded Green function,
$\mathbf{G}^r=\mathbf{G}^{0r}+\mathbf{G}^{0r}\mathbf{\S}^r\mathbf{G}^r$,
and a Keldysh equation for the lesser Green function
$\mathbf{G}^<=\mathbf{G}^r\mathbf{\Sigma}^<\mathbf{G}^a$, where
$\mathbf{G}^{0r}$ is the uncoupled dot Green function. In these
equations the self energies are the sum of the left and right self
energies, i.e., $\S^{(r,<)}=\S^{L(r,<)}+\S^{R(r,<)}$. A self
consistent calculation is required to calculate $\langle
n_{\overline{i}}\rangle$ and $\langle d_{\overline{i}}^\dagger d_i
\rangle$, which are given by the lesser Green function, $\langle
d_{j}^\dagger d_i\rangle=\int \frac{d\w}{2\pi}\mathrm{Im}
G_{ij}^<(\w)$.

{\it Noise.} The current operator can be written as its average
value plus some fluctuation, i.e., $\widehat{I}_\eta=I_\eta+\d
\widehat{I}_\eta$. In our system there are two sources of noise,
namely, thermal noise and shot noise. The first one is due to
thermal fluctuations in the occupations of the leads. It vanishes
for zero temperature and $eV\neq 0$, but can be finite for $T\neq
0$ and $eV=0$. On the other hand, shot noise is due to the
granularity of the electron charge and is a nonequilibrium
property of the system in the sense that it is nonzero only when
there is a finite current ($eV\neq 0$). To calculate the noise
(thermal+shot noise) we use the standard definition
$S_{\eta\eta'}(t-t')=\langle\{\d\widehat{I}_\eta
(t),\d\widehat{I}_{\eta'}(t')\}\rangle$, which can also be written
as $S_{\eta\eta'}(t-t')=\langle
\{\widehat{I}_\eta(t),\widehat{I}_{\eta'}(t')\}\rangle-2I_\eta^2$.
After a straightforward calculation, which will be presented
elsewhere \cite{fms02}, we find for the noise power spectrum ({\it
dc} limit) \cite{bd02}
\begin{equation}\label{}
\begin{split}
  S_{\eta\eta'}(0)=\frac{e^2}{\hbar}\int \frac{d\w}{2\pi} &{\rm Tr}\{
    \d_{\eta\eta'}in_\eta \mathbf{\G}^\eta \mathbf{G}^>
    -\d_{\eta\eta'}i(1-n_\eta)\mathbf{\G}^\eta \mathbf{G}^<+\mathbf{G}^<\mathbf{\G}^\eta \mathbf{G}^>\mathbf{\G}^{\eta'}\\&
    -n_\eta(1-n_{\eta'})\mathbf{G}^r\mathbf{\G}^\eta \mathbf{G}^r
    \mathbf{\G}^{\eta'}-n_{\eta'}(1-n_\eta)\mathbf{G}^a \mathbf{\G}^\eta \mathbf{G}^a \mathbf{\G}^{\eta'}\\&
   - \mathbf{G}^<\mathbf{\G}^\eta [(1-n_{\eta'})\mathbf{G}^r-(1-n_\eta)\mathbf{G}^a]\mathbf{\G}^{\eta'}+(n_\eta
    \mathbf{G}^r-n_{\eta'}\mathbf{G}^a)\mathbf{\G}^\eta \mathbf{G}^>
    \mathbf{\G}^{\eta'}\}.
\end{split}
\label{Setaeta}
\end{equation}
The $dc$ noise (zero frequency) is position independent, and it is
possible to show that $S_{LL}(0)=S_{RR}(0)=-S_{LR}(0)=-S_{RL}(0)$
\cite{ymb00}. In the next section we use the component $S_{LL}$.

\section{Results.} Using Eqs. (\ref{ILfinal}) and (\ref{Setaeta}) we calculate current and
noise for the system in Fig.1. We assume $\G_\s^\eta$ to be
independent of energy, but polarization dependent with values
$\G^L_\s=\G_0[1+(-1)^{\d_{\s\downarrow}}p]$,
$\G^R_\uparrow=\G^L_\uparrow$ and
$\G^R_\downarrow=\G^L_\downarrow$ if the leads have parallel
alignment or $\G^R_\downarrow=\G^L_\uparrow$ and
$\G^R_\uparrow=\G^L_\downarrow$ if they are antiparallel aligned.
The parameter $p$ gives the spin-splitting of the ferromagnetic
band. For example, for $p=0$ the system is unpolarized while for
$p=1$ the system is fully polarized. The parameter $\G_0$ fixes
the coupling strength between leads and dot. The sign +/- in
$\G_\s^L$ corresponds to majority/minority spins, respectively.
Here we take $\s=\uparrow$ ($\s=\downarrow$) as majority
(minority) spins in the lead $L$ and assume $\G_0=0.01U$ and
$p=0.5$ as in Ref.\cite{wr01}. The majority/minority spin
population in the right lead switches from one to the other
according to the lead polarization, which can be controlled via an
external magnetic field. This simple form for $\G_\s^\eta$ is
reasonable when the band is wide compared to others energies of
the system. The temperature is assumed to be $k_BT=\G_0(1+p)$. Our
approximation (Hartree-Fock) does not include correlations of the
Kondo type, however we do not expect these to change our results
in the present range of parameters.

\begin{figure}[h]
\begin{center}
\includegraphics[width=7cm]{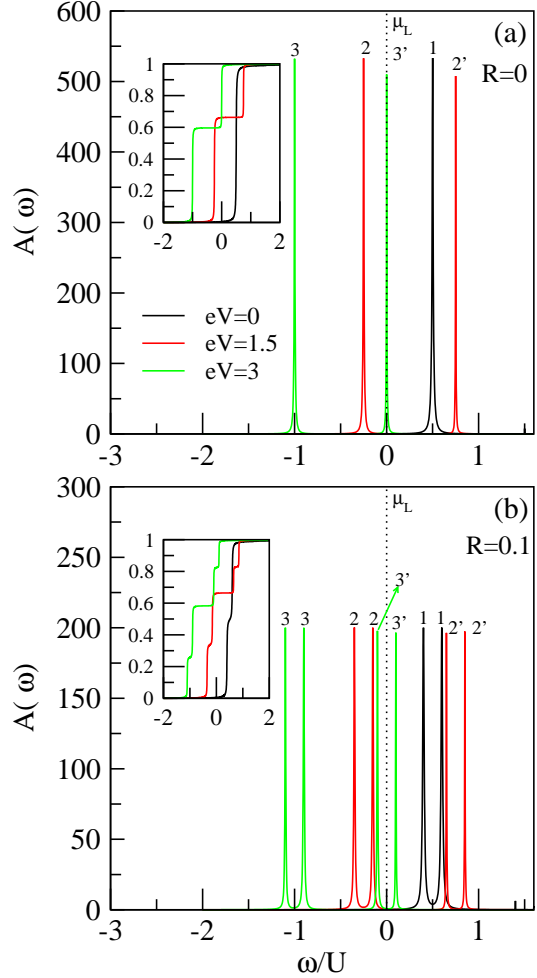}
\end{center}
\caption{Spectral function $A(\w)$ against energy $\w$ for $R=0$
and $R=0.1U$. The peaks correspond to the dot levels. For $R=0$
there is one peak for $eV=0$ (peak 1) and two peaks for $eV=1.5U$
and $eV=3U$ (peaks 2,2' and 3,3'). The extra peak (2' or 3') is
due to Coulomb interaction, since the lowest level (2 or 3) is
already below the Fermi energy (here at zero), thus allowing
electrons to go in the dot. For $R=0.1U$ the peaks are split and
given by $\e_0\pm R$ and $\e_0 \pm R+U$. The insets show the
integral $I(\w)$ of the spectral function. Each step gives the
area under a peak. Since the total area is normalized the last
step is at one.} \label{fig2}
\end{figure}

A relevant quantity in transport is the spectral function, for the
present spin-dependent case defined as
$A(\w)=i\mathrm{Tr}[\mathbf{G}^r(\w)-\mathbf{G}^a(\w)]$, whose
poles give the resonant levels which work as conduction channels.
Figure 2 shows $A(\w)$ for different applied voltages and for
$R=0$ (upper panel) or $R=0.1$ (lower panel). For $R=0$ we have
only one peak when $eV=0$ (labelled 1) and two peaks when
$eV=1.5U$ or $eV=3U$ (labelled 2,2' or 3,3', respectively). When
$eV=0$ the dot is empty because the level $\e_0=\e_{d}=0.5U$ is
above the Fermi energies $\mu_L$ and $\mu_R$ (set equal to zero),
so Coulomb interaction plays no role. When $eV=1.5U$ or $3U$,
$\e_0$ is below the Fermi energy of the left lead (peaks 2 or 3),
consequently the electrons could go inside the dot, creating the
high energy peak at $\e_0+U$ (peaks 2' or 3'), due to Coulomb
interaction. The levels in the dot shift linearly with the bias,
following the assumption $\e_0=\e_{d}-\frac{eV}{2}$. As mentioned
above, this linear drop does not account for charging effects.
However, it gives reasonable qualitative results here. For
$R=0.1U$ we have similar behaviors but each peak in the $R=0$ case
is now split due to spin-flip. The peaks are located at $\e_1$,
$\e_2$, $\e_1+U$ and $\e_2+U$ [Fig.\ref{fig2}(b)].

The inset in Fig.\ref{fig2}(a) shows the integrals of the spectral
function, namely $I(\w)=\frac{1}{4\pi}\int_{-\infty}^\w
A(\w')d\w'$ for the three voltages used. Observe that
$I(\w)\rightarrow 1$ as $\w$ increases. This is due to the
normalization of the spectral function. For $eV=0$ the whole area
is essentially under the peak at $0.5U$ (peak 1), which explains
the single step in $I(\w)$. For $eV=1.5U$ and $eV=3U$ the total
normalized area should be distributed under the two peaks (2 and
2' or 3 and 3'), in order to keep the normalization of $A(\w)$. It
leads to a reduction of the area of the lowest peak (2 or 3) in
comparison to its $eV=0$ value. This area is given by the first
step in $I(\w)$.

\begin{figure}[h]
\begin{center}
\includegraphics[width=7cm]{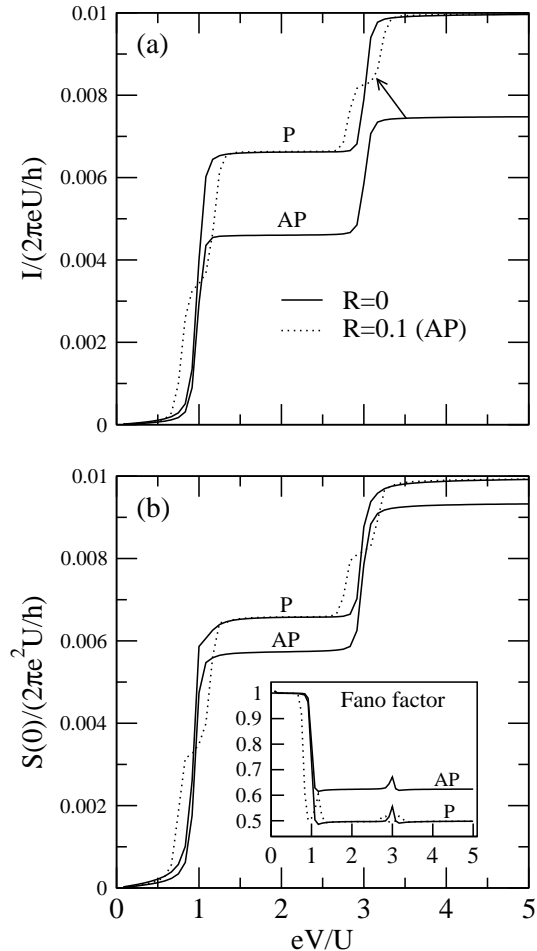}
\end{center}
\caption{Current and noise as a function of the bias for parallel
(P) and antiparallel (AP) alignments and with $R=0$ and $0.1U$.
The curves for $R=0.1U$ are only for the AP alignment; observe
that these are almost on top of the P curves, except within the
sloping region around $U$ and $3U$. Both current and noise are
reduced when the right lead changes its polarization from P to AP,
following the typical behavior of TMR. The inset shows a
suppression of the AP-Fano factor due to spin-flip.} \label{fig4}
\end{figure}

Figure \ref{fig4} shows current (a) and noise (b) as a function of
the bias with $R=0$ (solid line) and $R=0.1U$ (dotted line) for
both P and AP configurations. Because P and AP curves for $R=0.1U$
coincide, we plotted only the AP case. The first enhancement of
the current and noise at $eV=U$ happens when $\e_0$ crosses the
left chemical potential, allowing electrons to tunnel from the
emitter (left lead) to the dot and then to the collector (right
lead). The current and noise remain constant until the second
level $\e_0+U$ reaches $\mu_L$ at $eV=3U$, when another
enhancement is observed. In terms of differential conductance
($\s_{\rm diff}$) each enhancement corresponds to a peak in
$\s_{\rm diff}$. These peaks reflect the spectral function plotted
in Fig.2.

When the system changes from parallel (P) to antiparallel (AP)
configurations the current is reduced. This is a typical behavior
of tunnelling magnetoresistance (TMR): the resistance increases
when the system switches from P to AP configuration. The noise is
also affected by this resistance variation, showing a similar
reduction. Contrasting behaviors between current and noise will be
explored elsewhere \cite{fms02} for another set of parameters.

Looking at the effects of spin-flip on current and noise we see
that the AP curves with $R=0.1U$ (dotted lines) tend to be on the
P curves with $R=0$, thus showing that lead alignments are less
important when spin-flip plays a part. This AP current enhancement
due to spin-flip gives rise to a reduction of the TMR; since
$TMR=(I_P-I_{AP})/I_{AP}$, when $I_{AP}\rightarrow I_P$ we have
$TMR\rightarrow 0$. W. Rudzi{\'n}ski {\it et al.}\cite{wr01} found
a similar behavior for TMR.

In the inset of Fig.\ref{fig4} we plot the Fano factor
$S_{LL}/2eI_L$. For the parallel case the Fano factor remains
around 0.5 for voltages between $U$ and $5U$, except at $eV=3U$
where it has a small peak. This average value around $0.5$ is a
consequence of the symmetry of the double-barrier structure in the
P case. A similar behavior is observed for the AP case with its
average value above the P case. When spin-flip is included
($R=0.1U$) the AP Fano factor is shifted down, becoming close to
the P result for $R=0$, with the addition of a peak close to
$eV=U$ and a double peak around $eV=3U$. This peculiar double
structure is a consequence of the splitting of the dot levels when
$R\neq 0$ as observed in the spectral function.

\newpage

\section{Conclusion} Using the Keldysh nonequilibrium technique we
calculated current and noise in a ferromagnetic-quantum
dot-ferromagnetic system with Coulomb interaction and spin-flip
relaxation. We have shown that the lead alignments affect both
current and noise. These are reduced when the leads rotate from
the P to the AP configuration, following the typical
magnetoresistance behavior. The spin-flip relaxation is crucial to
drive the current and noise in the AP case close to their values
in the P case. In a way, we can say that spin-flip makes the P and
the AP configurations ``degenerate" thus reducing the effect of
the lead-polarization alignment on transport. We also showed that
TMR is reduced due to spin-flip, corroborating previous results in
the literature.

FMS acknowledges support from the funding agencies CAPES and
FAPESP (Brazil). JCE acknowledges financial support from the Swiss
NSF, DARPA and ARO.

\vspace{30 truecm}



\begin{thebibliography}{99}
\bibitem{commentRHO} The density matrix is defined as $\rho=\frac{e^{-\beta h}}{\mathrm{Tr}\{e^{-\beta
h}\}}$, where $h=H_L+H_R+H_D$. The coupling Hamiltonian is not
included in this definition because it is considered the
nonequilibrium part of the Hamiltoninan which is turned on at
$t=-\infty$. For futher discussion about this definition and its
implications we refer the rearder to \cite{hh96} and references
therein.
\end{thebibliography}
\end{document}